
\input harvmac.tex
\Title{CTP/TAMU-35/92}{{A Heterotic Multimonopole Solution}
\footnote{$^\dagger$}{Work supported in part by NSF grant PHY-9106593.}}

\centerline{
Ramzi~R.~ Khuri\footnote{$^*$}{Supported by a World Laboratory Fellowship.}}
\bigskip\centerline{Center for Theoretical Physics}
\centerline{Texas A\&M University}\centerline{College Station, TX 77843}

\vskip .3in
An exact multimonopole solution of heterotic string theory
is presented. The solution is constructed by a modification of the
't Hooft ansatz for a four-dimensional instanton. An analogous solution in
Yang-Mills field theory saturates a Bogomoln'yi bound and possesses the
topology and far field limit of a multimonopole configuration, but has
divergent action near each source. In the string solution, however, the
divergences from the Yang-Mills sector are precisely cancelled by those from
the gravity sector. The resultant action is finite and easily computed. The
Manton metric on moduli space for the scattering of two string monopoles
is found to be flat to leading order in the impact parameter, in agreement
with the trivial scattering predicted by a test monopole calculation.

\Date{4/92}

\def\sqr#1#2{{\vbox{\hrule height.#2pt\hbox{\vrule width
.#2pt height#1pt \kern#1pt\vrule width.#2pt}\hrule height.#2pt}}}
\def\Box{\mathchoice\sqr64\sqr64\sqr{4.2}3\sqr33}

\def\hstar {H^\ast_\mu}
\def\met {g_{\mu\nu}}

\def\b#1{\vec\beta_{#1}}

\lref\reyone {S.--J.~Rey {\it Axionic String Instantons
and their Low-Energy Implications}, Proceedings, Tuscaloosa 1989,
Superstrings and particle theory, p.291.}

\lref\reytwo {S.--J.~Rey, Phys. Rev. {\bf D43} (1991) 526.}

\lref\abenone{I.~Antoniadis, C.~Bachas, J.~Ellis and D.~V.~Nanopoulos,
Phys. Lett. {\bf B211} (1988) 393.}

\lref\abentwo{I.~Antoniadis, C.~Bachas, J.~Ellis and D.~V.~Nanopoulos,
Nucl. Phys. {\bf B328} (1989) 117.}

\lref\mtone{R.~R.~Metsaev and A.~A.~Tseytlin, Phys. Lett.
{\bf B191} (1987) 354.}

\lref\mttwo{R.~R.~Metsaev and A.~A.~Tseytlin,
Nucl. Phys. {\bf B293} (1987) 385.}

\lref\cfmp{C.~G.~Callan, D.~Friedan, E.~J.~Martinec
and M.~J.~Perry, Nucl. Phys. {\bf B262} (1985) 593.}

\lref\ckp{C.~G.~Callan,
I.~R.~Klebanov and M.~J.~Perry, Nucl. Phys. {\bf B278} (1986) 78.}

\lref\love{C.~Lovelace, Phys. Lett. {\bf B135} (1984) 75.}

\lref\fridven{B.~E.~Fridling and A.~E.~M.~Van de Ven,
Nucl. Phys. {\bf B268} (1986) 719.}

\lref\gepwit{D.~Gepner and E.~Witten, Nucl. Phys. {\bf B278} (1986) 493.}

\lref\quartet{D.~J.~Gross,
J.~A.~Harvey, E.~J.~Martinec and R.~Rohm, Nucl. Phys. {\bf B267} (1986) 75.}

\lref\dine{M.~Dine, Lectures delivered at
TASI 1988, Brown University (1988) 653.}

\lref\brone{E.~A.~Bergshoeff and M.~de Roo, Nucl.
Phys. {\bf B328} (1989) 439.}

\lref\brtwo{E.~A.~Bergshoeff and M.~de Roo, Phys. Lett. {\bf B218} (1989) 210.}

\lref\chsone{C.~G.~Callan, J.~A.~Harvey and A.~Strominger, Nucl. Phys.
{\bf B359} (1991) 611.}

\lref\chstwo{C.~G.~Callan, J.~A.~Harvey and A.~Strominger, Nucl. Phys.
{\bf B367} (1991) 60.}

\lref\bpst{A.~A.~Belavin, A.~M.~Polyakov, A.~S.~Schwartz and Yu.~S.~Tyupkin,
Phys. Lett. {\bf B59} (1975) 85.}

\lref\thooft{G.~'t~Hooft, Nucl. Phys. {\bf B79} (1974) 276.}

\lref\hoofan{G.~'t~Hooft, Phys. Rev. Lett., {\bf 37} (1976) 8.}

\lref\wil{F.~Wilczek, in {\it Quark confinement and field theory},
Ed. D.~Stump and D.~Weingarten, John Wiley and Sons, New York (1977).}

\lref\cofa{E.~Corrigan and D.~B.~Fairlie, Phys. Lett. {\bf B67} (1977) 69.}

\lref\jackone{R.~Jackiw, C.~Nohl and C.~Rebbi, Phys. Rev. {\bf D15} (1977)
1642.}

\lref\jacktwo{R.~Jackiw, C.~Nohl and C.~Rebbi, in {\it Particles and
Fields}, Ed. David Boal and A.~N.~Kamal, Plenum Publishing Co., New York
(1978), p.199.}

\lref\rkinst{R.~R.~Khuri, Phys. Lett. {\bf B259} (1991) 261.}

\lref\rkscat{C.~G.~Callan and R.~R.~Khuri, Phys. Lett. {\bf B261} (1991) 363.}

\lref\rkmant{R.~R.~Khuri, {\it Manton Scattering of String Solitons}
PUPT-1270 (to appear in Nucl. Phys. {\bf B}).}

\lref\rkdg{R.~R.~Khuri, {\it Some Instanton Solutions in String
Theory} to appear in Proceedings of the XXth International Conference on
Differential Geometric Methods in Theoretical Physics, World Scientific,
October 1991.}

\lref\rkthes{R.~R.~Khuri, {\it Solitons and Instantons in String Theory},
 Princeton University Doctoral Thesis, August 1991.}

\lref\rksing{M.~J.~Duff, R.~R.~Khuri and J.~X.~Lu, {\it String and
Fivebrane Solitons: Singular or Non-singular?}, Texas A\&M preprint,
CTP/TAMU-89/91 (to appear in Nucl. Phys. {\bf B}).}

\lref\rkorb{R.~R.~Khuri and H.~S.~La, {\it Orbits of a String around a
Fivebrane}, Texas A\&M preprint, CTP/TAMU-95/91
(submitted to Phys. Rev. Lett.).}

\lref\rkmot{R.~R.~Khuri and H.~S.~La, {\it String Motion in Fivebrane
Geometry}, Texas A\&M preprint, CTP/TAMU-98/91 (submitted to Nucl. Phys. B).}

\lref\rkmonex{R.~R.~Khuri {\it A Heterotic Multimonopole Solution},
Texas A\&M preprint, CTP/TAMU-35/92.}

\lref\rkmono{R.~R.~Khuri {\it A Multimonopole Solution in
String Theory}, Texas A\&M preprint, CTP/TAMU-33/92.}

\lref\rkmscat{R.~R.~Khuri {\it Scattering of String Monopoles},
Texas A\&M preprint, CTP/TAMU-34/92.}

\lref\ginsp{P.~Ginsparg, Lectures delivered at
Les Houches summer session, June 28--August 5, 1988.}

\lref\swzw {W.~Boucher, D.~Friedan and A.~Kent, Phys. Lett.
{\bf B172} (1986) 316.}

\lref\dghrr{A.~Dabholkar, G.~Gibbons, J.~A.~Harvey and F.~Ruiz Ruiz,
Nucl. Phys. {\bf B340} (1990) 33.}

\lref\dabhar{A.~Dabholkar and J.~A.~Harvey,
Phys. Rev. Lett. {\bf 63} (1989) 478.}

\lref\prso{M.~K.~Prasad and C.~M.~Sommerfield, Phys. Rev. Lett. {\bf 35}
(1975) 760.}

\lref\jim{J.~A.~Harvey and J.~Liu, Phys. Lett. {\bf B268} (1991) 40.}

\lref\mantone{N.~S.~Manton, Nucl. Phys. {\bf B126} (1977) 525.}

\lref\manttwo{N.~S.~Manton, Phys. Lett. {\bf B110} (1982) 54.}

\lref\mantthree{N.~S.~Manton, Phys. Lett. {\bf B154} (1985) 397.}

\lref\atiyah{M.~F.~Atiyah and N.~J.~Hitchin, Phys. Lett. {\bf A107}
(1985) 21.}

\lref\atiyahbook{M.~F.~Atiyah and N.~J.~Hitchin, {\it The Geometry and
Dynamics of Magnetic Monopoles}, Princeton University Press, 1988.}

\lref\strom{A.~Strominger, Nucl. Phys. {\bf B343} (1990) 167.}

\lref\gsw{M.~B.~Green, J.~H.~Schwartz and E.~Witten,
{\it Superstring Theory} vol. 1, Cambridge University Press (1987).}

\lref\polch{J.~Polchinski, Phys. Lett. {\bf B209} (1988) 252.}

\lref\dfluone{M.~J.~Duff and J.~X.~Lu, Nucl. Phys. {\bf B354} (1991) 141.}

\lref\dflutwo{M.~J.~Duff and J.~X.~Lu, Nucl. Phys. {\bf B354} (1991) 129.}

\lref\dfluthree{M.~J.~Duff and J.~X.~Lu, Phys. Rev. Lett. {\bf 66}
(1991) 1402.}

\lref\dflufour{M.~J.~Duff and J.~X.~Lu, Nucl. Phys. {\bf B357} (1991)
534.}

\lref\dfstel{M.~J.~Duff and K.~S.~Stelle, Phys. Lett. {\bf B253} (1991)
113.}

\lref\ferone{R.~C.~Ferrell and D.~M.~Eardley, Phys. Rev. Lett. {\bf 59}
(1987) 1617.}

\lref\fertwo{R.~C.~Ferrell and D.~M.~Eardley, {\it Slowly Moving
Maximally Charged Black Holes} in Frontiers in Numerical Relativity,
Cambridge University Press, 1987.}

\lref\gh{G.~W.~Gibbons and S.~W.~Hawking, Phys. Rev. {\bf D15}
(1977) 2752.}

\lref\ghp{G.~W.~Gibbons, S.~W.~Hawking and M.~J.~Perry, Nucl. Phys. {\bf B318}
(1978) 141.}

\lref\briho{D.~Brill and G.~T.~Horowitz, Phys. Lett. {\bf B262} (1991)
437.}

\lref\gidone{S.~B.~Giddings and A.~Strominger, Nucl. Phys. {\bf B306}
(1988) 890.}

\lref\gidtwo{S.~B.~Giddings and A.~Strominger, Phys. Lett. {\bf B230}
(1989) 46.}

\lref\raj{R.~Rajaraman, {\it Solitons and Instantons}, North Holland,
1982.}

\lref\chsw{P.~Candelas, G.~T.~Horowitz, A.~Strominger and E.~Witten,
Nucl. Phys. {\bf B258} (1984) 46.}

\lref\bogo{E.~B.~Bogomolnyi, Sov. J. Nucl. Phys. {\bf 24} (1976) 449.}

\lref\cogo{E.~Corrigan and P.~Goddard, Comm. Math. Phys. {\bf 80} (1981)
575.}

\lref\wardone{R.~S.~Ward, Comm. Math. Phys. {\bf 79} (1981) 317.}

\lref\wardtwo{R.~S.~Ward, Comm. Math. Phys. {\bf 80} (1981) 563.}

\lref\wardthree{R.~S.~Ward, Phys. Lett. {\bf B158} (1985) 424.}

\lref\groper{D.~J.~Gross and M.~J.~Perry, Nucl. Phys. {\bf B226} (1983)
29.}

\lref\ash{{\it New Perspectives in Canonical Gravity}, ed. A.~Ashtekar,
Bibliopolis, 1988.}

\lref\lich{A.~Lichnerowicz, {\it Th\' eories Relativistes de la
Gravitation et de l'Electro-magnetisme}, (Masson, Paris 1955).}

\lref\goldstein{H.~Goldstein, {\it Classical Mechanics}, Addison-Wesley,
1981.}

\lref\dflufive{M.~J.~Duff and J.~X.~Lu, Class. Quant. Grav. {\bf 9}
(1992) 1.}

\lref\dflusix{M.~J.~Duff and J.~X.~Lu, Phys. Lett. {\bf B273} (1991)
409.}

\lref\hlp{J.~Hughes, J.~Liu and J.~Polchinski, Phys. Lett. {\bf B180}
(1986).}

\lref\town{P.~K.~Townsend, Phys. Lett. {\bf B202} (1988) 53.}

\lref\duff{M.~J.~Duff, Class. Quant. Grav. {\bf 5} (1988).}

\lref\rossi{P.~Rossi, Physics Reports, 86(6) 317-362.}

\lref\gksone{B.~Grossman, T.~W.~Kephart and J.~D.~Stasheff, Commun. Math.
Phys. {\bf 96} (1984) 431.}

\lref\gkstwo{B.~Grossman, T.~W.~Kephart and J.~D.~Stasheff, Commun. Math.
Phys. {\bf 100} (1985) 311.}

\lref\gksthree{B.~Grossman, T.~W.~Kephart and J.~D.~Stasheff, Phys. Lett.
{\bf B220} (1989) 431.}

\newsec{Introduction}

In recent work several classical solitonic solutions of string theory with
higher-membrane structure have been presented. In \rkinst, the
tree-level axionic instanton solution of \reyone\ is extended to an exact
solution of bosonic string theory for the special case of a linear dilaton
\refs{\abenone,\abentwo} wormhole solution. Exactness is shown by combining
the metric and antisymmetric tensor in a generalized curvature
\refs{\mtone,\mttwo}, which
is written covariantly in terms of the tree-level dilaton field, and rescaling
the dilaton order by order in the parameter $\alpha'$. An exact heterotic
multi-soliton solution with instanton structure in the four dimensional
transverse space can be obtained\refs{\rkdg,\chsone,\chstwo} by equating
the curvature of the Yang-Mills gauge field with the above generalized
curvature. This latter solution represents an exact extension of the tree-level
fivebrane solutions of \refs{\dfluone,\dflutwo}.

In this paper we present an exact heterotic multi-soliton solution
which represents a multimonopole configuration. We obtain this
solution via a modification of the 't Hooft ansatz for the Yang-Mills
instanton. We identify an analogous multimonopole solution in field theory
with divergent action and indicate how in the string solution these
divergences are cancelled. We also study the dynamics of the string
monopoles and find that, unlike BPS monopoles, the string monopoles
scatter trivially to leading order in the impact parameter.

We first review in section 2 the basic bosonic solution with monopole-like
structure discussed in \rkthes. A tree-level multi-soliton solution for the
massless fields of the string is written. The corresponding single source
wormhole solution is extended to order $\alpha'$. This latter solution is
noted to contain the basic outline of a stringy correction to a magnetic
monopole. We then summarize the tree-level monopole solution in $N=4$
supersymmetric low-energy string theory of \jim.

We proceed in section 3 to construct an exact heterotic multimonopole
solution by modifying the 't Hooft
ansatz\refs{\hoofan\wil\cofa\jackone{--}\jacktwo} for the Yang-Mills instanton.
We note the relationship of this solution to the exact multi-instanton solution
in \chsone. Unlike the latter solution, however, the multimonopole solution
does not lend itself easily to a CFT description.

We note in section 4 that an analogous field theory solution representing
a multimonopole configuration not in the Prasad-Sommerfield\prso\ limit can
be immediately obtained from the modified 't Hooft ansatz independently of
string theory. This solution has the topology of $Q=1$ monopole sources,
saturates the Bogomoln'yi bound\bogo\ and exhibits the far field behaviour of
multimonopole sources. However, the action for this solution diverges near
each source.

We demonstrate in section 5 that the string solution, by contrast, has finite
action. The divergences coming from the Yang-Mills sector are precisely
cancelled by those from the gravitational sector. The resultant action
reduces to the tree-level form and is easily calculated. The zero force
condition for string solitons is seen to arise as a direct result
of the force cancellation in the gauge sector, once the generalized connection
and gauge connection are identified.

In section 6 we study the scattering of two string monopoles by
two methods. The first approach computes the Manton metric on moduli
space, which defines distance on the static solution manifold. A flat
metric is obtained to leading order in the impact parameter. This result
is consistent with a calculation of the dynamic force on a test string
monopole moving in the background of a source string monopole.

We conclude in section 7 with a discussion of our results and their
implications.

\newsec{Bosonic and Tree-Level Solutions}

In this section we briefly review two previously obtained solutions:
the bosonic multi-soliton solution obtained in \rkthes\ and the
Prasad-Sommerfield monopole\prso\ solution to supersymmetric low-energy
superstring theory in \jim. Both classes of solutions possess
three-dimensional spherical symmetry, as opposed to the four-dimensional
spherical symmetry of other instanton and fivebrane
solutions\refs{\rkinst,\strom,\dfluone,\dflutwo,\chsone,\chstwo}.

The tree-level bosonic multi-soliton solution to the string equations of
motion is given by\rkthes\
\eqn\bomons{\eqalign{e^{2\phi}&=C+\sum_{i=1}^N
{m_i\over |\vec x -\vec a_i|},\cr
g_{\mu\nu}&=e^{2\phi}\delta_{\mu\nu},\qquad\qquad \mu,\nu=1,2,3,4,\cr
g_{ab}&=\eta_{ab},\qquad\qquad a,b=0,5,6...25,\cr
H_{\alpha\beta\gamma}&=\pm\epsilon_{\alpha\beta\gamma}{}^\mu\partial_\mu
\phi,\qquad \alpha,\beta,\gamma,\mu=1,2,3,4,\cr}}
where $\phi$ is the dilaton, $g_{MN}$ is the string sigma model metric
and $H_{MNP}=\partial_{[M}B_{NP]}$, where $B_{NP}$ is the antisymmetric
tensor. $\vec x=(x_1,x_2,x_3)$ is a three-dimensional coordinate
vector in the $(123)$ subspace of the four-dimensional transverse space
$(1234)$. $m_i$ represents the charge and $a_i$ the location in the
three-space of the $i$th source.

Note that we have singled out a direction $x_4$ and projected out all
the field dependence on $x_4$. By doing so, we destroy the $SO(4)$ invariance
in the transverse space possessed by the instanton solution\rkinst. However,
\bomons\ is an equally valid solution to the string equations as the
multi-instanton solution with $e^{2\phi}=1+\sum_{i=1}^N
{Q_i\over |\vec x -\vec a_i|^2}$, where in this case the vectors are
four-dimensional, since in both cases the dilaton field satisfies the
Poisson equation $e^{-2\phi}\Box\ e^{2\phi}=0$. The projection is necessary
to obtain the three-dimensional symmetry of a magnetic monopole.

Although the above bosonic multi-soliton solution \bomons\ lacks the gauge
and Higgs fields normally attributed to a magnetic monopole in field theory,
one can think of the dual field in the transverse four-space
$\hstar \equiv {1\over 6}\epsilon_{\alpha\beta\gamma\mu}
H^{\alpha\beta\gamma}$ as the magnetic field strength of a multimonopole
configuration in the space $(123)$ (note that $H^\ast_4=0$).

Since the dilaton equation is essentially unaffected when we try to
obtain a tree-level supersymmetric solution, we can follow the derivation
of Duff and Lu's fivebrane solution\dfluone, but assume that the fields
are independent of one coordinate (say $x_4$), and again obtain a $D=10$
multi-fivebrane solution which breaks half the spacetime supersymmetries,
but with monopole-like structure.

Unlike the four-dimensional (instanton) solutions, the three-dimensional
solutions do not easily lend themselves to a CFT description, and it is
therefore difficult to go beyond $O(\alpha')$ in obtaining stringy
corrections to the tree-level fields. In \rkinst, the $O(\alpha')$ correction
was worked out for the
special case of a single source with $C=0$. The metric and antisymmetric
tensor were unchanged to $O(\alpha')$, but the dilaton is corrected:
\eqn\bomona{e^{2\phi}={m\over r}\left(1-{\alpha' \over 8mr}\right).}
Note that, unlike the $O(\alpha')$ correction to the four-dimensional
solution in \rkinst, the dilaton correction is not a simple rescaling of
the power of $r$ to order $\alpha'$. This fact is intimately connected
with the difficulty in formulating a CFT description of the three-dimensional
solution.

We now briefly summarize the tree-level monopole solution of \jim.
Starting with $N=1, D=10$ supergravity coupled to super Yang-Mills,
Harvey and Liu find a solution to the equations of motion with background
fermi fields set to zero. Supersymmetry requires that there exists
a positive chirality Majorana-Weyl spinor $\epsilon$ satisfying

\eqn\suei{\delta\psi_M=\left(\nabla_M-{\textstyle {1\over 4}}H_{MAB}\Gamma^{AB}
\right)\epsilon=0,}
\eqn\sueii{\delta\lambda=\left(\Gamma^A\partial_A\phi-{\textstyle{1\over 6}}
H_{AMC}\Gamma^{ABC}\right)\epsilon=0,}
\eqn\sueiii{\delta\chi=F_{AB}\Gamma^{AB}\epsilon=0,}
where $\psi_M,\ \lambda$ and $\chi$ are the gravitino, dilatino and gaugino
fields. The Bianchi identity is given by
\eqn\bianchi{dH=\alpha' \left(\tr R\wedge R-{\textstyle{1\over 30}}\Tr
        F\wedge F\right).}
Choose the spacetime indices to be $0,1,2,3$ and the internal indices to
be $4,5...9$. The $(9+1)$-dimensional Majorana-Weyl fermions decompose down to
chiral spinors according to $SO(9,1)\supset SO(3,1)\otimes SO(6)$ for
the $M^{9,1}\to M^{3,1}\times M^6$ decomposition. Again if we single out
a direction in internal space (say $x_4$), the above supersymmetry
equations and Bianchi identity are solved by a constant chiral
spinor\jim\ $\epsilon_\pm = \pm \Gamma^{1234}\epsilon_\pm$ and the ansatz
\eqn\harliu{\eqalign{F_{\mu\nu}=&\pm{1\over 2}\epsilon_{\mu\nu}
{}^{\lambda\sigma}F_{\lambda\sigma},\cr
H_{\mu\nu\lambda}=&\mp\epsilon_{\mu\nu\lambda}{}^\sigma\partial_\sigma\phi,\cr
g_{MN}=&{\rm diag}(-1,e^{2\phi},e^{2\phi},e^{2\phi},e^{2\phi},1,1,1,1,1),\cr
\nabla_\rho\nabla^\rho=&\mp {1\over 4} \alpha'\epsilon^{\mu\nu\lambda\sigma}
{\rm tr}F_{\mu\nu}F_{\lambda\sigma},\cr}}
where $\mu,\nu,\lambda,\sigma=1,2,3,4$.
The BPS monopole solution for the gauge and Higgs fields is given
by\refs{\prso,\bogo}
\eqn\bps{\eqalign{A_i^a&=\epsilon_{iab}{x^b\over r^2}(K-1),\cr
\Phi^a&={x^a\over r^2}H,\cr}}
where $H=Cr \coth{Cr}-1$, $K={Cr\over \sinh{Cr}}$ and $C$ is the
vacuum expectation value of the Higgs. Making the identification
$A_4^a\equiv \Phi^a$, replacing \bps\ into \harliu\
and solving the dilaton equation yields
\eqn\hldil{e^{2\phi}=e^{2\phi_0}+2\alpha'{1\over r^2}\left[ 1-K^2+2H
\right],}
which is nonsingular at $r=0$ and represents a single monopole source.

Since \harliu\ can be solved by any (anti) self-dual configuration, we can
in principle write down a multimonopole solution. While this solution is
supersymmetric, it is only tree-level in $\alpha'$, and not necessarily an
exact solution ({\it i.e.} in principle, we would have to obtain corrections
to the fields to higher order in $\alpha'$).

\newsec{Exact Heterotic Multimonopole Solution}

In this section we construct an exact multimonopole solution of heterotic
string theory. The derivation of this solution closely parallels that
of the multi-instanton solution presented in \refs{\chsone,\chstwo}, but
in this case, the solution possesses three-dimensional spherical symmetry near
each source, which turns out to represent a magnetic monopole of topological
charge $Q=1$. Again the reduction is effected by singling out a direction in
the
transverse space.

The supersymmetry equations \suei, \sueii\ and \sueiii\ are unchanged at
tree-level in heterotic string theory. In this case, however, the
$(9+1)$-dimensional Majorana-Weyl fermions decompose down to
chiral spinors according to $SO(9,1)\supset SO(5,1) \otimes SO(4)$ for
the $M^{9,1}\to M^{5,1}\times M^4$ decomposition.
Let $\mu,\nu,\lambda,\sigma=1,2,3,4$ and $a,b=0,5,6,7,8,9$. Then the ansatz
\eqn\anstz{\eqalign{\met&=e^{2\phi}\delta_{\mu\nu},\cr g_{ab}&=\eta_{ab},\cr
H_{\mu\nu\lambda}&=\pm\epsilon_{\mu\nu\lambda\sigma}\partial^\sigma\phi\cr}}
with constant chiral spinors $\epsilon_\pm$ again solves the supersymmetry
equations (again with zero background fermi fields) provided the YM gauge
field satisfies the instanton (anti)self-duality condition
\eqn\yminst{F_{\mu\nu}=\pm {1\over 2}\epsilon_{\mu\nu}{}^{\lambda\sigma}
F_{\lambda\sigma}.}
An exact solution is obtained as follows. Define a generalized connection by
\eqn\genc{\Omega^{AB}_{\pm M}=\omega^{AB}_M\pm H^{AB}_M}
embedded in an SU(2) subgroup of the gauge group, and equate it
to the gauge connection $A_\mu$\dine\ so that $dH=0$ and the corresponding
curvature $R(\Omega_{\pm})$ cancels against the Yang-Mills field strength $F$.
The crucial point is that for $e^{-2\phi}\Box\ e^{2\phi}=0$ with the
above ansatz, the curvature of the generalized connection can be written in the
covariant form\rkinst
\eqn\gencurv{\eqalign{R(\Omega_\pm)_{\mu\nu}^{mn}
=&\delta_{n\nu}\nabla_m\nabla_\mu\phi
- \delta_{n\mu}\nabla_m\nabla_\nu\phi + \delta_{m\mu}\nabla_n\nabla_\nu\phi
- \delta_{m\nu}\nabla_n\nabla_\mu\phi \cr
&\pm \epsilon_{\mu mn\alpha}\nabla_\alpha\nabla_\nu\phi
\mp \epsilon_{\nu mn\alpha}\nabla_\alpha\nabla_\mu\phi ,\cr}}
from which it easily follows that
\eqn\gcinst{R(\Omega_\pm)^{mn}_{\mu\nu}=
\mp\half\epsilon_{\mu\nu}^{\ \ \ \lambda\sigma}
R(\Omega_{\pm})_{\lambda\sigma}^{mn}.}
Thus we have a solution with the ansatz \anstz\ such that
\eqn\exsol{F_{\mu\nu}^{mn}=R(\Omega_{\pm})_{\mu\nu}^{mn},}
where both $F$ and $R$ are (anti)self-dual.
This solution becomes exact since $A_\mu=\Omega_{\pm\mu}$
implies that all the higher order corrections vanish\refs{\brone,\brtwo}.
The self-dual solution for the gauge connection is then given by the 't Hooft
ansatz for the four-dimensional instanton
\eqn\hfanstz{A_\mu=i \overline{\Sigma}_{\mu\nu}\partial_\nu \ln f,}
where $\overline{\Sigma}_{\mu\nu}=\overline{\eta}^{i\mu\nu}(\sigma^i/2)$
for $i=1,2,3$ ($\sigma^i$, $i=1,2,3$ are the $2\times 2$ Pauli matrices), where
\eqn\hfeta{\eqalign{\overline{\eta}^{i\mu\nu}=-\overline{\eta}^{i\nu\mu}
&=\epsilon^{i\mu\nu},\qquad\qquad \mu,\nu=1,2,3,\cr
&=-\delta^{i\mu},\qquad\qquad \nu=4 \cr}}
and where $f^{-1}\Box\ f=0$. The ansatz for the anti-self-dual solution
is similar, with the $\delta$-term in \hfeta\ changing sign.

To obtain a multi-instanton solution, one solves for $f$ in the
four-dimensional space to obtain
\eqn\finst{f=e^{-2\phi_0}e^{2\phi}=1+\sum_{i=1}^N{\rho_i^2\over
|\vec x - \vec a_i|^2},}
where $\rho_i^2$ is the instanton scale size and $\vec a_i$ the location in
four-space of the $i$th instanton.

To obtain a multimonopole solution, we modify the 't Hooft ansatz as follows.
We again single out a direction
in the transverse four-space (say $x_4$) and assume all fields are independent
of this coordinate. Then the solution for $f$ can be written as
\eqn\fdmono{f=e^{-2\phi_0}e^{2\phi}=1+\sum_{i=1}^N{m_i\over
|\vec x - \vec a_i|},}
where $m_i$ is the charge and $\vec a_i$ the location in
the three-space $(123)$ of the $i$th source. If we make the identification
$\Phi\equiv A_4$, then the gauge and Higgs fields may be simply written
in terms of the dilaton as
\eqn\stmono{\eqalign{\Phi^a&=-{2\over g}\delta^{ia}\partial_i\phi,\cr
A_k^a&=-{2\over g}\epsilon^{akj}\partial_j\phi\cr}}
for the self-dual solution. For the anti-self-dual solution, the Higgs
field simply changes sign. Here $g$ is the YM coupling constant. Note that
$\phi_0$ drops out in \stmono.
The solution in \fdmono\ can be thought of as a multi-line source instanton
solution, each monopole being interpreted as an ``instanton string''\rossi.

The above solution (with the gravitational fields obtained
directly from \anstz\ and \fdmono) represents an exact multimonopole
solution of heterotic string theory. In order to more clearly see the
monopole structure of this solution, we first consider in the next section
an analogous solution in field theory and study its properties, which
then carry over directly into the string solution.

\newsec{Multimonopole Solution in Field Theory}

We now turn to an analogous multimonopole solution in field theory.
Consider the four-dimensional Euclidean action
\eqn\ymact{S=-{1\over 2g^2}\int d^4x {\rm Tr} G_{\mu\nu}G^{\mu\nu},
\qquad\qquad \mu,\nu =1,2,3,4.}
For gauge group $SU(2)$, the fields may be written as $A_\mu=(g/2i)
\sigma^a A_\mu^a$ and $G_{\mu\nu}=(g/2i)\sigma^a G_{\mu\nu}^a$.
The equation of motion derived from this action is solved by the
modified 't Hooft ansatz shown in the previous section:
\eqn\hfanstz{A_\mu=i \overline{\Sigma}_{\mu\nu}\partial_\nu \ln f,}
where again
\eqn\fmono{f=1+\sum_{i=1}^N{m_i\over |\vec x - \vec a_i|},}
where $m_i$ is the charge and $\vec a_i$ the location in the three-space
$(123)$ of the $i$th source.
To obtain a multimonopole solution, we again identify the scalar
field $\Phi\equiv A_4$ (we loosely refer to this field as a Higgs field in
this paper, although there is no apparent symmetry breaking mechanism).
The Lagrangian density for the above ansatz can be rewritten as
\eqn\lgdn{\eqalign{G_{\mu\nu}^a G_{\mu\nu}^a =&G_{ij}^a G_{ij}^a +
2G_{k4}^a G_{k4}^a \cr =&G_{ij}^a G_{ij}^a + 2D_k \Phi^a D_k \Phi^a , \cr}}
which has the same form as the Lagrangian density for YM + massless scalar
field in three dimensions.

We now go to $3+1$ dimensions with the Lagrangian density (signature $(-+++)$)
\eqn\ymhlag{{\cal L}=-{1\over 4}G_{\mu\nu}^a G^{\mu\nu a} -{1\over 2}
D_\mu \Phi^a D^\mu \Phi^a,}
and show that the above multi-soliton ansatz is a static solution with
$A_0^a=0$
and all time derivatives vanish. The equations of motion in this
limit are given by
\eqn\ymheqs{\eqalign{D_i G^{jia}&=g\epsilon^{abc}(D^j\Phi^b)\Phi^c,\cr
D_i D^i \Phi^a&=0.\cr}}
It is then straightforward to verify that the above equations are solved by
\eqn\monsol{\eqalign{\Phi^a&=\mp{1\over g}\delta^{ai}\partial_i \omega,\cr
A_k^a&=\epsilon^{akj}\partial_j \omega,\cr}}
where $\omega\equiv \ln f$. This solution represents a multimonopole
configuration with sources at $\vec a_i=1,2...N$. A simple observation of
far field and near field behaviour shows that this solution does not arise
in the Prasad-Sommerfield\prso\ limit. In particular, the fields are singular
near the sources and vanish as $r\to\infty$.

The topological charge of each source is easily computed
($\hat\Phi^a\equiv {\Phi^a/|\Phi|}$) to be
\eqn\topo{Q=\int d^3x k_0={1\over 8\pi}\int d^3x\epsilon_{ijk}\epsilon^{abc}
\partial_i\hat\Phi^a\partial_j\hat\Phi^b\partial_k\hat\Phi^c=1.}
The magnetic charge of each source is then given by $m_i=Q/g=1/g$.
It is also straightforward to show that the Bogomoln'yi\bogo\ bound
\eqn\bobo{G_{ij}^a=\epsilon_{ijk}D_k\Phi^a}
is saturated by this solution. Finally, it is easy to show that
the magnetic field $B_i={1\over 2}\epsilon_{ijk}F^{jk}$ (where
$F_{\mu\nu}\equiv \hat\Phi^a G_{\mu\nu}^a-(1/g)\epsilon^{abc}\hat\Phi^a
D_\mu \hat\Phi^b D_\nu \hat\Phi^c$ is the gauge-invariant electromagnetic
field tensor defined by 't Hooft\thooft) has the the far field limit behaviour
of a multimonopole configuration:
\eqn\bfield{B(\vec x)\to \sum_{i=1}^N {m_i(\vec x - \vec a_i)\over
|\vec x - \vec a_i|^3},\qquad {\rm as}\quad r\to \infty.}
As usual, the existence
of this static multimonopole solution owes to the cancellation of the
gauge and Higgs forces of exchange--the ``zero-force'' condition.

We have presented all the monopole properties of this solution.
Unfortunately, this solution as it stands has divergent
action near each source, and this singularity cannot be simply removed
by a unitary gauge transformation. This can be seen for a single source
by noting that as $r\to 0$, $A_k\to {1\over 2}\left(U^{-1}\partial_k U\right)$,
where $U$ is a unitary $2\times 2$ matrix. The expression in parentheses
represents a pure gauge, and there is no way to get around the $1/2$ factor in
attempting to ``gauge away'' the singularity\raj. The field theory
solution is therefore not very interesting physically. As we shall see in the
next section, however, the string theory solution has far greater potential.

\newsec{Finiteness of String Solution}

The string solution presented in section 3 has the same structure in
the four-dimensional transverse space as the multimonopole solution
of the YM + scalar field action of section 4. If we identify
the $(123)$ subspace of the transverse space as the space part of the
four-dimensional spacetime (with some toroidal compactification, similar
to that used in \jim) and take the timelike direction as the usual $X^0$,
then the monopole properties described in the previous section carry
over directly into the string solution.

The string action contains a term $-\alpha' F^2$ which also diverges
as in the field theory solution. This divergence, however,
is precisely cancelled by the term $\alpha' R^2(\Omega_\pm)$ in the
$O(\alpha')$
action. This result follows from the exactness condition
$A_\mu=\Omega_{\pm\mu}$
which leads to $dH=0$ and the vanishing of all higher order corrections
in $\alpha'$. Another way of seeing this is to consider the higher order
corrections to the bosonic action shown in \refs{\brone,\brtwo}. All such
terms contain the tensor $T_{MNPQ}$, a generalized curvature incorporating
both $R(\Omega_\pm)$ and $F$. The ansatz is contructed precisely so that this
tensor vanishes identically\refs{\rkinst,\rkdg}. The action thus reduces to
its lowest order form and can be calculated directly for a multi-source
solution from the expressions for the massless fields in the gravity sector.

The divergences in the gravitational sector in heterotic string theory thus
serve to cancel the divergences stemming from the field theory solution. This
solution thus provides an interesting example of how this type of cancellation
can occur in string theory, and supports the promise of string theory as a
finite theory of quantum gravity. Another point of interest is that the string
solution represents a supersymmetric multimonopole solution coupled to gravity,
in which the zero-force condition in the gravitational sector ({\it i.e.}
the cancellation between the attractive gravitational force and repulsive
antisymmetric tensor force) arises as a direct result of the zero-force
condition in the gauge sector (cancellation between gauge and Higgs exchange
forces) once the gauge connection and generalized connection are identified.

We now calculate the mass of the heterotic multimonopole configuration.
Naively, the mass can be calculated
from the tree-level action (since the higer order terms drop out)
\eqn\volaction{S=-{1\over 2\kappa^2}\int d^3x\sqrt{g} e^{-2\phi}\left( R
+ 4(\nabla\phi)^2 - {H^2\over 12}\right).}
There is one subtlety we must consider, however (see \rkmant).
 From the term $\sqrt{g} e^{-2\phi} R$ in the integrand of the action,
the action density in \volaction\ contains double derivative terms of the
metric component fields. In general, one would like to work with an action
which depends only on the fields and their first derivatives. This
problem was solved in general relativity by Gibbons and Hawking\refs
{\gh,\ghp}, who added a surface term which precisely cancelled the
double derivative terms in the action in general relativity.
The addition of a surface term does not, of course, affect the equations
of motion.

It turns out that there is a relatively straightforward generalization
of the Gibbons-Hawking surface term (GHST) to string
theory\refs{\gidone,\briho}.
By antisymmetry, the axion field does not contribute to the GHST and
the surface term in this case can be written in the simple form
\eqn\stghst{S_{GHST}=-{1\over \kappa^2}\int_{\partial M}\left(e^{-2\phi}K-
K_0\right),}
where $\partial M$ is the surface boundary and $K$ and $K_0$ are the
traces of the fundamental form of the boundary surface embedded in the
metric $g$ and the Minkowskian metric $\eta$ respectively. The correct
effective action is thus obtained by adding the surface term of \stghst\
to the volume term of \volaction:
\eqn\trueaction{S=-{1\over 2\kappa^2}\left[\int d^3x
\sqrt{g} e^{-2\phi}
\left( R + 4(\nabla\phi)^2 - {H^2\over 12}\right)
+2\int_{\partial M}\left(e^{-2\phi}K-K_0\right)\right].}

By using the equations of motion, the volume term $S_V$ can
be written as a surface term (see \rkmant):
\eqn\svoltwo{S_V=-{1\over \kappa^2}\int_{\partial M} \hat
n\cdot\vec \nabla e^{-2\phi}.}
Note that $\sqrt{g}$ has been absorbed into the surface measure of
$\partial M$. Since we have separability of sources in the limit of surfaces of
infinite radius, we may therefore compute $S_V$ for a single monopole
configuration in three-space
\eqn\simono{\eqalign{e^{2\phi}&=1+{m\over r},\cr
g_{ij}&=e^{2\phi}\delta_{ij},\cr}}
and simply add the contributions of an arbitrary number of sources.
The contribution of a single monopole to the static volume action is given by
\eqn\mvol{\eqalign{S_V&=-{1\over \kappa^2}
({\partial\over \partial r}e^{-2\phi})A(M)\cr
&=-{4\pi m\over \kappa^2}\cr}}
in the $r\to\infty$ limit,
where $A(M)=4\pi r^2(1+m/r)$ is the area of the boundary surface.

We now turn to the GHST.
A simple calculation of the extrinsic curvature $K$ for a single monopole
configuration \simono\ gives
\eqn\kcurved{K={2\over r^2}e^{-3\phi}(r+m/2).}
When the surface $\partial M$ is embedded in flat space, the radius of
curvature $R$ is given by $R=re^\phi$. The extrinsic curvature $K_0$ is
then given by
\eqn\kflat{K_0={2\over R}={2\over r}e^{-\phi}.}
The GHST is therefore  given by
\eqn\ghstone{S_{GHST}=-{2\over
\kappa^2r}\left(e^{-5\phi}( 1+ {m\over 2r}) - e^{-\phi}\right) A(M)=
{12\pi m\over \kappa^2}}
in the $r\to\infty$ limit.

The total static action for a multi-soliton configuration, equal to the
total mass of the solitons, can then be obtained by adding the static
contributions to the action of the volume part and the GHST. The result is
\eqn\totmass{M_T={8\pi\over \kappa^2}\sum_{n=1}^N m_n.}
For our multimonopole configuration, however, it should be noted that
$m_n=1/g$ for $n=1,2...N$.

\newsec{Dynamics of String Monopoles}

We now consider the dynamics of the string monopoles.
For this purpose, we adopt two different methods. The first
computes the Manton metric on moduli space for the scattering of
two string monopoles, while the second studies the motion of a test string
monopole in the background of a source string monopole. We will
find that the two methods yield consistent results.

Manton's prescription\manttwo\ for the study of soliton scattering may be
summarized as follows. We first invert the constraint equations of the system.
The resultant time dependent field configuration does not in general
satisfy the full time dependent field equations, but provides
an initial data point for the fields and their time derivatives.
Another way of saying this is that the initial motion is tangent to the
set of exact static solutions.  The kinetic action obtained
by replacing the solution to the constraints into the action defines a
metric on the parameter space of static solutions. This metric defines
geodesic motion on the moduli space\manttwo.

A calculation of the metric on moduli space for the scattering of BPS
monopoles and a description of its geodesics was worked out by Atiyah
and Hitchin\atiyah. Several interesting properties of monopole
scattering were found, such as the conversion of monopoles into dyons
and the right angle scattering of two monopoles on a direct collision
course\refs{\atiyah,\atiyahbook}. The configuration space is found to
be a four-dimensional manifold $M_2$ with a self-dual Einstein metric.

In this section, we adapt Manton's prescription to study the
dynamics of heterotic string monopoles. A similar procedure was
followed in \rkmant\ for the Manton scattering of heterotic instantons.
Indeed, many of the formal computations carry over from the instanton
computation. For the monopoles, however, the divergences plagueing the
instanton calculation are absent, thus rendering our task far simpler.
In both cases, we follow essentially
the same steps that Manton outlined for monopole scattering, but take
into account the peculiar nature of the string effective action. Since
we work in the low-velocity limit, our kinematic analysis is nonrelativistic.

We first solve the constraint equations for the soliton solutions.
These equations are simply the $(0j)$
components of the equations of motion (see \refs{\rkinst,\rkmant})
\eqn\constraints{\eqalign{R_{0j}-{1\over
4}H^2_{0j}+2\nabla_0\nabla_j\phi&=0,\cr
-{1\over 2}\nabla_kH^k{}_{0j}+H_{0j}{}^k\partial_k\phi&=0.\cr}}
Note that we use the tree-level equations of motion, as the higher order
corrections in $\alpha'$ automatically vanish.
We wish to find an $O(\beta)$ solution to the above equations which
represents a quasi-static version of \anstz\ (i.e. a solution of
the form \anstz\ but with time dependent $\vec a_i$). In other words,
we would like to give each source an arbitrary transverse velocity
$\vec\beta_n$ in the $(123)$ subspace of the four-dimensional transverse space
and see what corrections to the fields are required by the
constraints. The vector $\vec a_n$ representing the position of source
$n$ in the three-space $(123)$ is given by
\eqn\aunty{\vec a_n(t)=\vec A_n + \vec\beta_nt,}
where $\vec A_n$ is the initial position of the $n$th source. Note that at
$t=0$ we have an exact static multi-soliton solution. Our solution to
the constraints will adjust our quasi-static approximation so that the
initial motion in the parameter space is tangent to the initial
exact solution at $t=0$.

The $O(\beta)$ solution to the constraints is given by
\eqn\orderbeta{\eqalign{e^{2\phi(\vec x,t)}&=1+\sum_{n=1}^N{m_n\over
|\vec x - \vec a_n(t)|},\cr g_{00}&=-1,\qquad g^{00}=-1,\qquad
g_{ij}=e^{2\phi}\delta_{ij},\qquad g^{ij}=e^{-2\phi}\delta_{ij},\cr
g_{0i}&=-\sum_{n=1}^N{m_n\vec\beta_n\cdot \hat x_i\over |\vec x - \vec
a_n(t)|},\qquad g^{0i}=e^{-2\phi}g_{0i},\cr
H_{ijk}&=\epsilon_{ijkm}\partial_m e^{2\phi},\cr
H_{0ij}&=\epsilon_{ijkm}\partial_m g_{0k}=\epsilon_{ijkm}\partial_k
\sum_{n=1}^N{m_n\vec\beta_n\cdot \hat x_m\over |\vec x - \vec a_n(t)|},\cr}}
where $i,j,k,m=1,2,3,4$, the $\vec a_n(t)$ are given by \aunty\ and we use a
flat space $\epsilon$-tensor. Note that $g_{00}$, $g_{ij}$ and $H_{ijk}$ are
unaffected to order $\beta$. Also note that we can interpret the solitons
as either line sources in the four-dimensional space $(1234)$ or point
sources in the three-dimensional subspace $(123)$.

The kinetic Lagrangian is obtained by replacing the expressions for the
fields in \orderbeta\ into \trueaction. Since \orderbeta\ is a solution to
order $\beta$, the leading order terms in the action (after the
quasi-static part) are of order $\beta^2$. In the volume term of the
action, $O(\beta)$ terms in the solution give $O(\beta^2)$ terms in the
kinetic action. As explained in \rkmant, the contribution of the GHST
to the kinetic action can be written in the form $m_s{\beta^2}/2$ for
each source, and the contributions of the sources can be simply
added. The GHST does not therefore play an important role in the dynamics
of the string monopoles, but merely serves to give the correct total mass.
Collecting all $O(\beta^2)$ terms in $S_V$ we
get the following kinetic Lagrangian density for the volume term:
\eqn\kinlag{\eqalign{{\cal L}_{kin}=-{1\over 2\kappa^2}\Biggl(
&4\dot \phi\vec M\cdot\vec \nabla\phi
-e^{-2\phi}\partial_iM_j\partial_iM_j
-e^{-2\phi}M_k\partial_j\phi\left(\partial_jM_k-\partial_kM_j\right)\cr
&+4M^2e^{-2\phi}(\vec \nabla\phi)^2
+2\partial_t^2e^{2\phi}-4\partial_t(\vec M\cdot\vec \nabla\phi)-4\vec
\nabla\cdot(\dot\phi\vec M)\Biggr),\cr}}
where $\vec M\equiv -\sum_{n=1}^N{m_n\vec \beta_n\over |\vec x - \vec
a_n(t)|}$. Henceforth let $\vec X_n\equiv \vec x - \vec a_n(t)$.
The last three terms in \kinlag\ are time-surface or space-surface terms
which vanish when integrated. Note that the kinetic Lagrangian has the
same form as in \rkmant. The contributions of the GHST are again simply
flat kinetic terms.

In contrast to the instanton case, the kinetic Lagrangian
$L_{kin}=\int d^3x{\cal L}_{kin}$ for monopole scattering converges
everywhere. This can be seen simply by studying the limiting behaviour
of $L_{kin}$ near each source. For a single source at $r=0$ with magnetic
charge $m$ and velocity $\beta$, we collect the logarithmically divergent
pieces
and find that they cancel:
\eqn\logdiv{{m\beta^2\over 2}\int r^2 drd\theta \sin\theta d\phi
\left(-{1\over r^3} + {3\cos^2\theta\over r^3}\right)=0.}
So unlike the instanton case, in which we were compelled to extract
information from the convergent interaction terms, in this case we can
use the self-terms directly.

We now specialize to the case of two heterotic monopoles of magnetic
charge $m_1=m_2=m=1/g$ and velocities $\vec\beta_1$ and $\vec\beta_2$.
Let the monopoles be located at $\vec a_1$ and $\vec a_2$.
Our moduli space consists of the configuration space of the relative
separation vector $\vec a\equiv \vec a_2 - \vec a_1$.
The most general kinetic Lagrangian can be written as
\eqn\genkinlag{\eqalign{L_{kin}=&h(a)(\b1\cdot\b1+\b2\cdot\b2)+p(a)\left(
(\b1\cdot\hat a)^2 + (\b2\cdot\hat a)^2\right)\cr
&+2f(a)\b1\cdot\b2 + 2g(a)(\b1\cdot\hat a)(\b2\cdot\hat a).\cr}}
Now suppose $\b1 = \b2 =\vec\beta$, so that \genkinlag\ reduces to
\eqn\boostlag{L_{kin}=(2h+2f)\beta^2+(2p+2g)(\vec\beta\cdot\hat a)^2.}
This configuration, however, represents the boosted solution of the
two-static soliton solution. The kinetic energy should therefore be
simply
\eqn\cmke{L_{kin}={M_T\over 2}\beta^2,}
where $M_T=M_1+M_2=2M={16\pi m}/{\kappa^2}$ is the total mass of
the two soliton solution. It then follows that the anisotropic part of
\boostlag\ vanishes and we have
\eqn\hfpg{\eqalign{g+p&=0,\cr 2(h+f)&={M_T\over 2}.\cr}}

It is therefore sufficient to compute $h$ and $p$. This can be done by
setting $\vec\beta_1=\vec\beta$ and $\vec\beta_2=0$.
The kinetic Lagrangian then reduces to
\eqn\rdkinlag{L_{kin}=h(a)\beta^2 + p(a)(\vec\beta\cdot\hat a)^2.}
Suppose for simplicity
also that $\vec a_1=0$ and $\vec a_2=\vec a$ at $t=0$.
The Lagrangian density of the volume term in this case is given by
\eqn\voltm{\eqalign{{\cal L}_{kin}&={-1\over 2\kappa^2}\Biggl(
{3m^3 e^{-4\phi}\over 2r^4}(\vec\beta\cdot\vec x)\left[
{\vec\beta\cdot\vec x\over r^3} + {\vec\beta\cdot(\vec x-\vec a)
\over |\vec x-\vec a|^3}\right] - {e^{-2\phi}m^2\beta^2\over r^4}\cr
&-{e^{-4\phi}m^3\beta^2\over 2r^4}\left( {1\over r} +
{\vec x\cdot(\vec x-\vec a)\over |\vec x-\vec a|^3}\right) +
{e^{-6\phi}m^4\beta^2\over r^2}\left( {1\over r^4} + {1\over |\vec x-\vec a|^4}
+ {2\vec x\cdot(\vec x-\vec a)\over r^3|\vec x-\vec a|^3}\right)\Biggr).\cr}}
The GHST contribution to the kinetic Lagrangian can be simply added
after integration and will not affect the analysis below.

The integration of the kinetic Lagrangian density in \voltm\ over three-space
yields the kinetic Lagrangian from which the metric on moduli space can be
read off. For large $a$, the nontrivial leading order  behaviour of the
components of the metric, and hence for the functions $h(a)$ and $p(a)$, is
generically of order $1/a$. In fact, for Manton scattering of YM monopoles,
the leading order scattering angle is $2/b$\mantthree, where $b$ is the impact
parameter. In this paper, we restrict our computation to the leading order
metric in moduli space. A tedious but straightforward collection of $1/a$
terms in the Lagrangian yields
\eqn\leadi{{-1\over 2\kappa^2}{1\over a}\int d^3x\left[ -{3m^4e^{-6\phi_1}
\over r^7}(\vec\beta\cdot\vec x)^2 + {m^3e^{-4\phi_1}\over r^4}\beta^2+
{m^4e^{-6\phi_1}\over r^5}\beta^2 - {3m^5e^{-8\phi_1}\over r^6}\beta^2
\right],}
where $e^{2\phi_1}\equiv 1+m/r$.
The first and third terms clearly cancel after integration over three-space.
The second and fourth terms are spherically symmetric. A simple integration
yields
\eqn\leadii{\int_0^\infty r^2dr \left( {e^{-4\phi_1}\over r^4} -
{3m^2e^{-8\phi_1}\over r^6}\right)
=\int_0^\infty {dr\over (r+m)^2} - 3m^2\int_0^\infty {dr\over (r+m)^4}=0.}
The $1/a$ terms therefore cancel, and the leading order metric on moduli
space is flat. This implies that the leading order scattering is trivial.
In other words, there is no deviation from the initial trajectories to
leading order in the impact parameter.

The above result is rather surprising and suggests that, in addition
to the static force, the leading order dynamic force also vanishes.
For pure YM monopoles, this is certainly not the case. For the string
monopoles, however, the dynamic YM force is precisely cancelled by the
dynamic gravity sector force.

To confirm this result, we employ the test-soliton approach of
\refs{\dghrr,\rkscat} to compute the dynamic force exerted on a
test string monopole moving in the background of a source string monopole.
Again only the massless fields in the gravitational sector come in to
play at tree-level. Since the monopoles have fivebrane structure, we
adopt the fivebrane action of Duff and Lu\refs{\dfluone,\dflutwo}
\eqn\sigfiv{\eqalign{S_{\sigma_5}=&-T_6\int d^6\xi\Biggl({1\over 2}
\sqrt{-\gamma}\gamma^{mn}\partial_mX^M\partial_nX^N g_{MN}e^{-\phi/6}
-2\sqrt{-\gamma}\cr&+{1\over 6!}\epsilon^{mnpqrs}\partial_mX^M\partial_nX^N
\partial_pX^P\partial_qX^Q\partial_rX^R\partial_sX^S A_{MNPQRS}\Biggr),\cr}}
where $m,n,p,q,r,s=0,5,6,7,8,9$ are fivebrane indices and
$M,N,P,Q,R,S=0,1,...9$ are spacetime indices
(transverse indices are denoted by $i,j=1,2,3,4$).
$\gamma_{mn}$ is a $5+1$-dimensional worldsheet metric, $g_{MN}$ is the
canonical spacetime metric and $A_{MNPQRS}$ is the
antisymmetric six-form potential whose curl $K=dA$ is
dual to the antisymmetric field strength $H_{\alpha\beta\gamma}$.

The multimonopole solution written in this frame is given by
\eqn\fivanstz{\eqalign{ds^2&=e^{2A}\eta_{mn}dx^mdx^n+e^{2B}\delta_{ij}
dx^idx^j, \cr  A_{056789}&=-e^C,\cr}}
where all other components of $A_{MNPQRS}$ are set to zero and the dilaton
$\phi$ and the scalar functions $A$, $B$ and $C$ are given by
\eqn\abc{\eqalign{A&=-{(\phi-\phi_0)\over 4},\cr
B&={3(\phi-\phi_0)\over 4},\cr C&=-2\phi+{3\phi_0\over 2},\cr}}
where $\phi_0$ is the value of the dilaton field at infinity and
\eqn\fivsltn{e^{2\phi}=e^{2\phi_0}\left(1+\sum_{n=1}^N
{m_n\over |\vec x -\vec a_n|}\right),}
where $\vec x$ and $\vec a_n$ are again vectors in the three-dimensional
subspace $(123)$ of the transverse space $(1234)$.

The Lagrangian for a test monopole moving in a background of identical static
source monopoles is given by substituting
\fivanstz\ in \sigfiv\ and then eliminating the worldbrane metric. The result
is
\eqn\fbworldsheet{{\cal L}_6=-T_6\left[\sqrt{-\det (e^{-2\phi/3+\phi_0/2}
\eta_{mn}+e^{4\phi/3-3\phi_0/2}\partial_m X^M\partial_n X_M)}
-e^{-2\phi+3\phi_0/2}\right].}

Since the test-monopole moves only in the $(123)$ subspace of the
transverse space (there is no motion along or field dependence on the
direction $x_4$), \fbworldsheet\ reduces in the low-velocity limit to
\eqn\fblagrange{\eqalign{{\cal L}_6&\simeq-T_6\left[e^{-2\phi+3\phi_0/2}
\left(1-\half e^{2(\phi-\phi_0)}(\dot X^i)^2\right)
-e^{-2\phi+3\phi_0/2}\right]\cr
&={T_6\over 2}e^{-\phi_0/2}(\dot X^i)^2~,\cr}}
where $i=1,2,3$.
Again both the static force and the nontrivial $O(v^2)$ velocity-dependent
force vanish. Hence this result also predicts trivial scattering, in direct
agreement with the flat Manton metric calculation.

\newsec{Conclusion}

In this paper, we have presented an exact multimonopole solution of
heterotic string theory. This solution represents a supersymmetric
extension of the bosonic string multimonopole solution outlined in
\rkthes, and is obtained by a modification of the 't Hooft
ansatz for a four-dimensional instanton. Exactness is shown by the
generalized curvature method used in \refs{\rkinst,\rkdg,\chsone,\chstwo}
to obtain exact instanton solutions in bosonic and heterotic string
theory. Unlike the instanton solutions, however, the monopole solutions
do not seem to be easily describable in terms of conformal field theories,
an unfortunate state of affairs from the point of view of string theory.

An analogous multimonopole solution of the four dimensional field theory
of YM + massless scalar field can be immediately written down. This solution
possesses the properties of a multimonopole solution (topology,
far-field limit and Bogomoln'yi bound) but has divergent action near
each source. In the string solution, however, these
divergences in the YM sector are cancelled by similar divergences in
the gravity sector, thus resulting in a finite action solution. This finding is
significant in that it represents an example of how string theory
incorporates gravity in such a way as to cancel infinities inherent in
gauge theories, thus supporting its promise as a theory of quantum gravity.

The cancellation between the gauge and gravitational sectors also
influences the dynamics of the string monopoles. Indeed, we find from
both a Manton metric on moduli space calculation and a test string
monopole calculation that the leading order dynamic force between two
string monopoles vanishes. This result implies trivial scattering between
string monopoles to leading order.

\vfil\eject
\listrefs
\bye